\documentclass[sigconf]{acmart}

\copyrightyear{2025} 
\acmYear{2025} 
\setcopyright{cc}
\setcctype{by}
\acmConference[SUI '25]{ACM Symposium on Spatial User Interaction}{November 10--11, 2025}{Montreal, QC, Canada}
\acmBooktitle{ACM Symposium on Spatial User Interaction (SUI '25), November 10--11, 2025, Montreal, QC, Canada}
\acmDOI{10.1145/3694907.3765933}
\acmISBN{979-8-4007-1259-3/2025/11}
\usepackage{xcolor}
\usepackage{tabularx}
\newcommand{\change}[1]{\textcolor{black}{#1}}

\newboolean{showRemovedSections}
\setboolean{showRemovedSections}{false}

\newcommand{\remove}[1]{}

\newcommand{\removeSection}[1]{%
  \ifthenelse{\boolean{showRemovedSections}}{#1}{}%
}

\newcommand{\revision}[1]{#1}

\AtBeginDocument{
  \providecommand\BibTeX{{%
    \normalfont B\kern-0.5em{\scshape i\kern-0.25em b}\kern-0.8em\TeX}}}

\begin{document}

\newcommand{\system}{RealitySummary}


\title[\system{}: Exploring On-Demand Mixed Reality \\ Text Summarization and Question Answering using Large Language Models]{\system{}: Exploring On-Demand Mixed Reality \\ Text Summarization and Question Answering using Large Language Models}

\author{Aditya Gunturu}
\affiliation{%
  \institution{University of Calgary}
  \city{Calgary}
  \country{Canada}}  
\affiliation{%
  \institution{University of Colorado Boulder}
  \city{Boulder}
  \country{USA}}  
\email{aditya.gunturu@ucalgary.ca}

\author{Shivesh Jadon}
\affiliation{%
  \institution{University of Calgary}
  \city{Calgary}
  \country{Canada}}  
\affiliation{%
  \institution{Apple}
  \city{Seattle}
  \country{USA}}    
\email{shivesh.jadon@ucalgary.ca}

\author{Nandi Zhang}
\affiliation{%
  \institution{University of Calgary}
  \city{Calgary}
  \country{Canada}}  
\affiliation{%
  \institution{University of Rochester}
  \city{Rochester}
  \country{USA}}  
\email{nandi.zhang@ucalgary.ca}

\author{Morteza Faraji}
\affiliation{%
  \institution{University of Calgary}
  \city{Calgary}
  \country{Canada}}  
\email{morteza.faraji@ucalgary.ca}

\author{Jarin Thundathil}
\affiliation{%
  \institution{University of Calgary}
  \city{Calgary}
  \country{Canada}}  
\email{jarin.thundathil@ucalgary.ca}


\author{Wesley Willett}
\affiliation{%
  \institution{University of Calgary}
  \city{Calgary}
  \country{Canada}}  
\email{wesley.willett@ucalgary.ca}

\author{Ryo Suzuki}
\affiliation{%
  \institution{University of Colorado Boulder}
  \city{Boulder}
  \country{USA}}
\email{ryo.suzuki@colorado.edu}

\renewcommand{\shortauthors}{Gunturu et al.}

\begin{abstract}
Large Language Models (LLMs) are gaining popularity as reading and summarization aids. However, little is known about their potential benefits when integrated with mixed reality (MR) interfaces to support everyday reading. In this iterative investigation, we developed RealitySummary \footnote{Project Page: \url{https://adigunturu.github.io/RealitySummary_SUI25/}}, an MR reading assistant that seamlessly integrates LLMs with always-on camera access, OCR-based text extraction, and augmented spatial and visual responses. Developed iteratively, RealitySummary evolved across three versions, each shaped by user feedback and reflective analysis: 1) a preliminary user study to understand reader perceptions (N=12), 2) an in-the-wild deployment to explore real-world usage (N=11), and 3) a diary study to capture insights from real-world work contexts (N=5). Our empirical studies' findings highlight the unique advantages of combining AI and MR, including always-on implicit assistance, long-term temporal history, minimal context switching, and spatial affordances, demonstrating significant potential for future LLM-MR interfaces beyond traditional screen-based interactions.
\end{abstract}



\begin{teaserfigure}
\includegraphics[width=\textwidth]{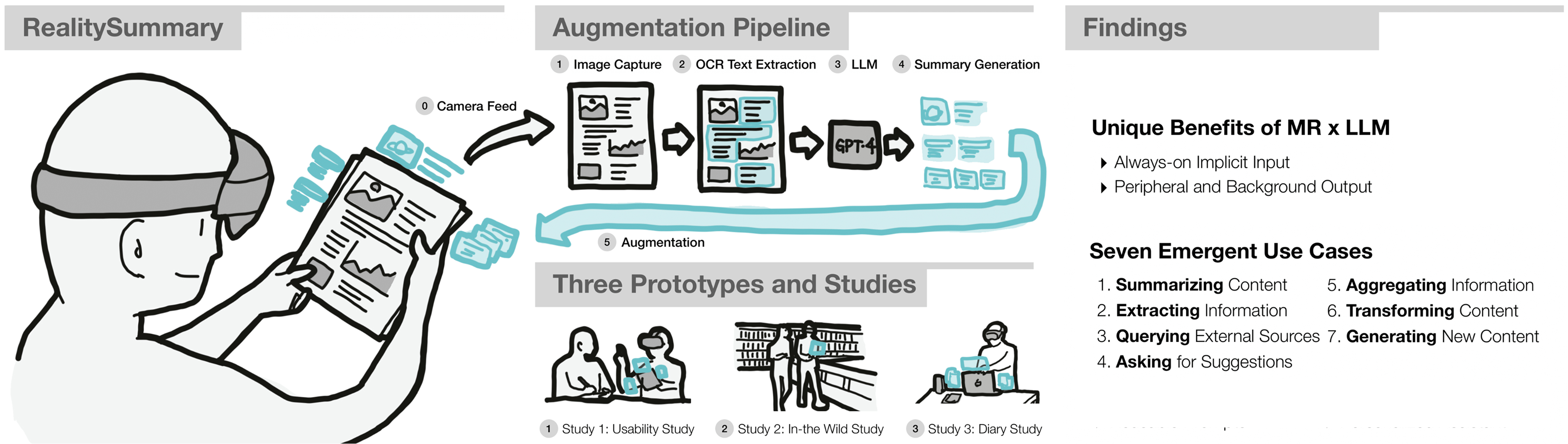}
\caption{RealitySummary is an on-demand mixed reality reading assistant developed over three iterations, with each version extensively shaped by user feedback and reflection. Through this evolution, we highlight key insights and two notable findings on integrating LLMs in MR interfaces: 1) the unique benefits of MR + LLMs, and 2) emergent use cases for MR reading assistants.}
\label{fig:teaser}
\end{teaserfigure}

\maketitle

\section{Introduction}


In his pioneering work~\cite{wellner1991digitaldesk}, Pierre Wellner introduced the concept of enhanced documents, advocating for a future where computer interfaces should augment our physical environment rather than restricting interactions to flat rectangle screens. Since then, HCI researchers have explored augmented reading experiences by leveraging Augmented Reality (AR) and Mixed Reality (MR) interfaces~\cite{li2019holodoc, rajaram2022paper, chen2020augmenting}. However, insights into the \textit{real-world, everyday use} of these interfaces remain largely unexplored. Most existing systems rely on pre-prepared information and preprocessed responses, limiting their applicability beyond controlled lab environments.

This paper provides empirical study contributions to understanding the benefits and limitations of mixed reality reading assistants in \textit{real-world} and \textit{everyday} contexts. We developed \system{}, an MR reading assistant powered by Large Language Models (LLMs) to ensure applicability and reliability in uncontrolled real-world environments. By leveraging an always-on camera, OCR-based text extraction, question-answering capabilities, and augmented spatial and visual LLM responses within MR interfaces, we \revision{deployed} this system beyond lab settings, uncovering unique insights.  \change{In this paper, we ask: How do real-time, always-on, augmented spatial and visual summaries in mixed reality complement reading experiences? Does this integration shape navigation and comprehension differently from traditional screen-based reading? What challenges and unique use cases emerge when deploying this in the wild? And what can be revealed about its potential long-term use?}


RealitySummary evolved through three iterative versions, each shaped by user feedback and reflection. First, we developed the initial prototype of \system{} (\textbf{Study 1: RS-Doc}) on the HoloLens 2 and conducted a preliminary user study (N=12) to understand reader perceptions and assess the usability of on-demand mixed reality reading assistants with AI-generated summarization. Based on the findings, we developed a second version of the system (\textbf{Study 2: RS-Wild}) on the Apple Vision Pro, enabling an in-the-wild study (N=11) to gather insights on real-world usage. Finally, we created the third iteration (\textbf{Study 3: RS-Diary}) to conduct a diary study (N=5), which explored how readers adapt to MR reading assistants over an extended period. \change{All three studies are presented in this paper as part of a single continuous investigation.}

These iterations and studies collectively highlight the unique benefits of integrating LLMs with MR for everyday assistance. For instance, the implicit input and spatial output modalities provide a clear advantage over traditional explicit inputs and screen-based outputs, allowing readers to stay focused on their tasks with minimal context switching. The always-on camera functionality also enables new, previously unconsidered use cases, such as enhancing not only documents but also posters, signage, nutrition labels, emails, and even restaurant menus in everyday environments with LLM enabled augmentations. Additionally, MR's spatial and tangible affordances open up novel ways to interact with LLMs, including summarizing entire bookshelves or multiple books on a table rather than just individual pages held in hand. These insights demonstrate how combined MR + LLM systems can effectively support long-term reading and knowledge work in daily activities.

Finally, our paper contributes:
\begin{enumerate} 
\item \system{}, an iteratively-developed mixed reality reading assistant that uses large language models for on-demand content summarization and augmentation. 
\item Observations and opportunities drawn from our usability study (N=12), in-the-wild study (N=11), and diary study (N=5), which highlight the potential benefits of mixed reality document enhancements. 
\item Lessons learned from the three iterations of our system. 
\end{enumerate}

\section{Related Work}
We summarize prior work on reading augmentation and content summarization with a focus on interactive approaches. 

\subsection{Reading Augmentation}
\subsubsection*{Document Enhancement}
Human-computer interaction researchers have explored ways to augment physical documents with AR-enhanced content. Examples include \textit{HoloDoc}~\cite{li2019holodoc}, \textit{Replicate and Resuse}~\cite{gupta2020replicate}, and \textit{PaperTrail}~\cite{rajaram2022paper}, all of which laid the groundwork for augmenting physical paper by superimposing dynamic virtual content onto printed documents. Prior work has demonstrated these concepts on different platforms. For example, \textit{Pacer}~\cite{liao2010pacer}, \textit{MagPad}~\cite{xu2015magpad}, \textit{Teachable Reality}~\cite{monteiro2023teachable}, and \textit{Dually Noted}~\cite{qian2022dually} use mobile AR for text augmentation, whereas \textit{Replicate and Resuse}~\cite{gupta2020replicate} and \textit{Wiki-TUI}~\cite{wu2007wikitui} leverage head-mounted displays to overlay information. Other examples such as \textit{DigitalDesk}~\cite{wellner1991digitaldesk}, \textit{DocuDesk}~\cite{everitt2008docudesk}, Matulic et al.'s pen and touch systems~\cite{matulic2013pen, matulic2013gesture}, and \textit{QOOK}~\cite{zhao2014qook} use projection mapping for document augmentation. \revision{Finally, \textit{cAR} \cite{hincapie2014car} leveraged transparent displays for document augmentation to support active reading.}
At their core, such augmented reading research~\cite{kirner2012development, billinghurst2001magicbook, hidayat2020magic, grasset2008design} aims to improve cognitive load~\cite{cheng2017reading}, knowledge accumulation~\cite{dunser2012creating, zhang2019design}, and spatial visualization~\cite{shelton2004exploring}, given a long-standing consensus that readers prefer physical paper over screen-based reading ~\cite{tashman2011active, sellen2003myth}. 

However, none of the existing systems have achieved \textit{on-demand} document enhancement. \revision{Here, we define \textit{on-demand} as the ability to dynamically and interactively enhance documents without requiring prior manual preparation or pre-processing.} \remove{Instead, documents required pre-processing with content manually prepared before being embedded in AR.} For example, prior work like \textit{HoloDoc}~\cite{li2019holodoc}, \textit{PaperTrail}~\cite{rajaram2022paper}, and \textit{Affinity Lens}~\cite{subramonyam2019affinity} did not extract text directly from the given document. Instead, content was prepared in advance and loaded based on attached fiducial markers. As a result, these systems are not fully adaptable and deployable to real-world scenarios. Some preliminary work has explored on-demand text analysis using machine learning, such as \textit{Dually Noted}~\cite{qian2022dually} analyzing document structure in real-time, \textit{Augmented Math}~\cite{chulpongsatorn2023augmented} and \textit{Augmented Physics}~\cite{gunturu2024augmented} extracting diagrams and math equations, and \textit{SOCRAR}~\cite{strecker2022socrar} using OCR to extract key information. However, none have yet achieved comprehensive and general-purpose document enhancement. 

\subsubsection*{Screen-Based Reading Support Tools}
Outside of the context of mixed reality, a wide range of projects and systems have explored reading support via screen-based interfaces. Examples include \textit{LiquidText}~\cite{tashman2011liquidtext}, \textit{texSketch}~\cite{subramonyam2020texsketch}, \textit{ScholarPhi}~\cite{head2021augmenting}, \textit{Chameleon}~\cite{masson2020chameleon},
and \textit{Marvista}~\cite{chen2022marvista}. Similar to our work, \textit{Marvista}~\cite{chen2022marvista} helps users read online documents by providing reading aids including summaries, contextual prompt questions, and reading metrics. Other research has studied active and close reading support, including \textit{Metatation}~\cite{mehta2017metatation} for close reading, \textit{XLibris}~\cite{price1998xlibris} for free form annotation, Matulic and Norrie's tabletop interfaces~\cite{matulic2012supporting} to support navigation, and \textit{GatherReader}~\cite{hinckley2012informal} for information gathering. 
Other systems have supported literature review and reference search for scientific papers~\cite{palani2023relatedly, kang2023comlittee, chang2023citesee}.
Lastly, works like \textit{Explorable Explanations}~\cite{victor2011explorable} and \textit{Potluck}~\cite{potluck} try to enhance static documents with interactive content to improve content comprehension through exploration. While this past work is limited to screen-based interfaces, our paper explores the augmented reading design for AR-based interfaces.

\subsection{Content Summarization}

\subsubsection*{Automatic Summarization}
In the field of natural language processing, a variety of tools have employed automatic summarization to support readers~\cite{el2021automatic, zhang2023concepteva}. These approaches provide various summarization methods, including abstractive and extractive summaries generated from document photos~\cite{benharrak2022summarylens}, real-time summarization of user writing for quick text iterations~\cite{dang2022beyond}, enhanced reading comprehension through automatic summarization~\cite{chen2022marvista}, and \textit{responsive text summaries} for adapting documents to screen sizes ranging from large displays to small watch faces~\cite{leiva2018responsive}.
With recent advances in large language models like GPT-4 and GPT-5, such automatic summarization tasks have become more accurate, robust, and accessible. For example, Goyal et al.~\cite{goyal2022news} showed that GPT-3 summaries were preferred by humans over those generated by fine-tuned models. Various types of summarization have been explored using GPT-3, such as opinion summarization~\cite{bhaskar2022zero}, news summarization~\cite{goyal2022news}, and medical dialogue summarization~\cite{chintagunta2021medically} while others have built custom frameworks for feeding long texts into GPT-3 to elicit more finely-controlled summarizations~\cite{Width_AI}. \newline

\subsubsection*{Interactive Summarization.} 
Researchers have also explored interactive summarization approaches based on context and users' needs.
For example, \textit{Semantic Reactor}~\cite{SemanticReactor}, developed by Google, enables users to ask questions about documents in natural language by fine-tuning BERT~\cite{devlin2019bert} using the SQuAD dataset~\cite{rajpurkar2018know}. \textit{VERSE}~\cite{vtyurina2019verse} is another Q\&A system that allows users to ask specific types of questions about a document but is limited by its inability to summarize. \textit{Hoeve et al.}~\cite{ter2020conversations}, meanwhile, identified a diverse questions asked by participants while reading documents, with factual/summarization questions being the most frequent. Other types include document-related, factoid, mechanical (answerable by rule-based methods), factual, yes/no, navigational, and summary questions. Voice assistants like Cortana, Alexa, Siri, VERSE~\cite{vtyurina2019verse}, Firefox Voice~\cite{cambre2021firefox}, and Pushpak~\cite{holani2019pushpak} offer efficient query modalities, as speech input has been found to be faster than manual input, especially for mixed reality devices~\cite{ruan2016speech, adhikary2021text}. 

\subsubsection*{\revision{Reality-Based Information Retrieval}}

\revision{Other recent research has examined visual reality-based information queries~\cite{buschel2018here}. For example, \textit{GazePointAR}~\cite{lee2024gazepointar} explored long term use of pronoun disambiguation in visual question answering in mixed-reality. \textit{G-Voila}~\cite{wang2024g} explored gaze as a medium for question answering in everyday scenarios.} \revision{While prior work has either focused on document augmentation with manual preparation \cite{li2019holodoc} or on general mixed-reality LLM-based QnA \cite{lee2024gazepointar}, our goal is to build on this foundation by developing and exploring on-demand and real-time reading augmentation within everyday Mixed-Reality settings.}

\change{To our knowledge, RealitySummary is the first LLM powered MR reading system to be studied in the wild over an extended period, enabling the collection of rich, longitudinal insights into real-world reading behaviors. This long-term, situated deployment serves as the foundation for our contribution, upon which we highlight implications of on-demand, context-aware mixed reality augmentations.}


\remove{The limitations of prior work such as Holodoc \cite{li2019holodoc}, as explained before, prevented the study of long term and in-the-wild use of MR reading assistants. To our knowledge GazePointAR \cite{lee2024gazepointar} is the closent work which explored long-term use of pronoun disambiguation in MR voice assistants. Still, it is unclear how can LLM powered MR assistants are used in everyday life to assist users in various types of reading and knowledge work tasks. We aim to fill this gap. In this paper, we contribute to the first demonstration of on-demand document enhancement by leveraging large language models. The goal of this paper is not only to demonstrate its potential but also to uncover its limitations and challenges from a user experience perspective through iterative implementation and multiple user evaluations.}
\begin{figure*}[h]
\includegraphics[width=\textwidth]{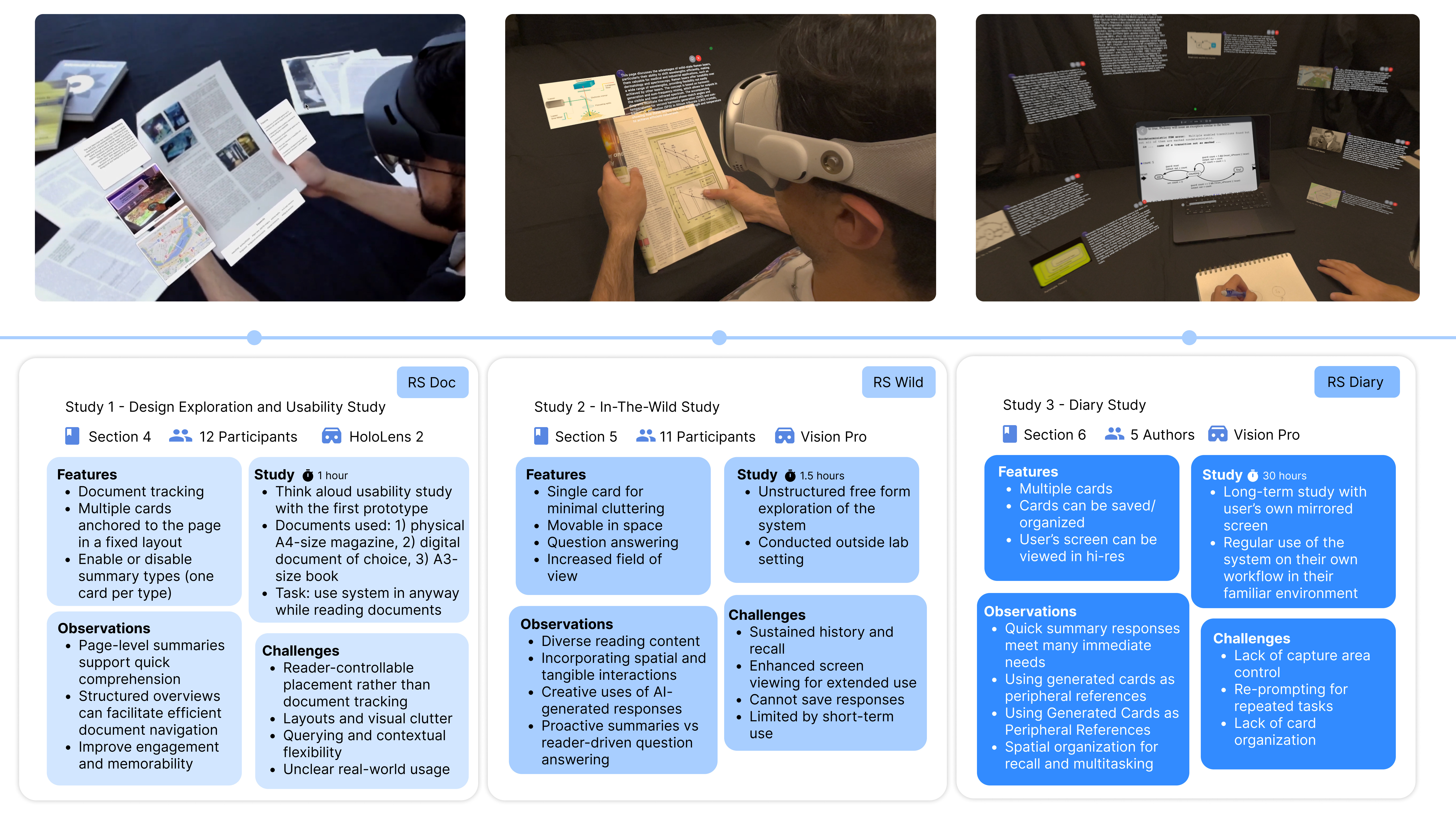}
\caption{Overview of our research approach spanning an evolution of our system across three different versions.}
\label{fig:system-design}
\end{figure*}
\section{Overview of Our Research Approach}

To uncover insights into on-demand reading assistants, we employed a three-phase iterative design process to investigate reader interactions and experiences with MR reading assistants. These explorations included a usability study using the initial prototype (Study~1: RS-Doc), followed by an in-the-wild study with a second prototype (Study~2: RS-Wild), and an author-conducted diary study with a third prototype (Study~3: RS-Diary). Each phase was designed to build upon the insights gained from the previous round, allowing for continuous refinement of the design and functionality.
\change{Insights from each study motivated the design of the subsequent system, and the three studies are presented together in this paper as part of a single continuous investigation, providing a complete picture of the iterative design process and cross-study findings.} \change{All studies were conducted under the approval of the University of Calgary ethics board (IRB Approval Number: REB21-2065). Participants provided informed consent prior to participation and could withdraw at any time without penalty.}

\remove{We chose this multi-method, iterative approach to capture a comprehensive understanding of user interactions across various contexts. The initial usability study provided controlled insights into the basic functionality and user experience of \system{}. The subsequent in-the-wild study allowed us to observe how users interacted with the improved, more reliable prototype in everyday environments, uncovering real-world usage patterns and challenges. Finally, the diary study offered deep, longitudinal insights into the integration of \system{} into daily life, allowing for nuanced observations of subtle interaction patterns and evolving user needs. This progression from controlled to naturalistic settings, coupled with iterative prototyping, allowed us to holistically explore the complex dynamics of AR-AI interactions while addressing emerging design considerations and ethical implications at each stage.}

\removeSection{
\section*{\remove{Study 0: Elicitation Study}}
}

\remove{Before implementing our system, we conducted a design elicitation study to explore potential designs for an AI-assisted augmented reading assistant. We held formative design workshops with six participants (4 males and 2 females) who had various design backgrounds as undergraduate and graduate HCI researchers. The study involved two tasks, each lasting approximately 15-20 minutes, with the entire session taking about 60 minutes.}

\removeSection{
\subsection*{\remove{Method}}
}

\removeSection{
\subsection*{\remove{Results}}
}

\section{RS-Doc: Design Exploration and Usability Study with the First Prototype}

We implemented the first version of our system (RS-Doc), which focused on in-place summarization for physical and digital documents, on a Microsoft HoloLens 2. In this section, we present initial observations from our experiences developing and evaluating RS-Doc as well as new design opportunities highlighted by those explorations.


\subsection{Implementation}



\subsubsection*{System Design}
RS-Doc captures documents using the HoloLens 2's internal camera and extracts text with Google Cloud OCR. The extracted text is processed using GPT-4 via OpenAI’s API to identify key information, while spaCy's named entity recognition (NER) module assists in identifying important entities. The system employs Vuforia for markerless image tracking, and the Mixed Reality Toolkit 3 (MRTK3) SDK for content rendering. Readers interact with the system through voice commands processed via Google's speech-to-text API, allowing them to modify or query visualizations. The system captures images periodically for OCR, with an option for manual activation, and tracks documents by generating dynamic image targets, ensuring accurate document tracking and augmented content rendering.

\subsubsection*{\revision{Enhancement Types}}
\revision{To design and implement our system, we identified five key categories of AI-generated content enhancements: \textbf{summarize}, \textbf{compare}, \textbf{augment}, \textbf{extract}, and \textbf{navigate}---based on an early elicitation study (\change{see appendix}) as well as prior work, such as \textit{HoloDoc}~\cite{li2019holodoc}, \textit{PaperTrail}~\cite{rajaram2022paper}, \textit{DuallyNoted}~\cite{qian2022dually}, \textit{QOOK}~\cite{zhao2014qook}, and \textit{Digital Desk}~\cite{li2019holodoc}. These categories informed both the conceptual framework and the system features. \textbf{Summarize} condenses text for quick reference, ranging from broad overviews to personalized summaries. This is implemented by sending the current page's text to GPT-4 for concise summaries, limited to 150 words.
\textbf{Compare} transforms data into visual formats by parsing documents. Our system provides the compare feature by generating \textit{comparison tables} for documents discussing multiple related themes, entities, or concepts, as well as \textit{timelines} for documents describing sequences of events.
\textbf{Augment} enhances content with external data. For instance, \textit{HoloDoc}~\cite{li2019holodoc} and \textit{PaperTrail}~\cite{rajaram2022paper} augment papers with external content, such as references, videos, and figures. Similarly, our system augments documents by providing \textit{information cards} about people and places using Google Search, Image APIs, and Mapbox, based on keywords identified by spaCy.
\textbf{Extract} enables readers to pull and store elements like keywords and quotes. For example, \textit{DigitalDesk}~\cite{wellner1991digitaldesk} allows users to extract content from documents for copy-and-paste functionality. RS-Doc implements this feature by generating \textit{keyword lists} that highlight the most central named entities extracted.
Finally, \textbf{Navigate} improves document navigation through features like a table of contents or bookmarks. For instance, \textit{QOOK}~\cite{zhao2014qook} allows users to add bookmarks to specific sections. RS-Doc supports navigation by generating a \textit{table of contents}, created by the LLM based on the document.}

\subsubsection*{Interactions}
\revision{After launching RS-Doc, a reader can trigger summaries for an in-view document by saying ``give me summaries''. The system then captures an image of the page currently in view, creates a dynamic image target, and generates summary cards of all types relevant to that page. Summaries are displayed around the page using a \textit{fixed layout} (for example, timeline summaries always appear at the top right of the page and information cards always appear to the left) and remain locked to the page if either the document or reader moves. 
For example, if a user is reading a magazine about aerial drones, the system automatically detects important keywords such as people, places, and phrases from the page. The system then generates an image, a comparison table, a summary, and information about key people and keywords. Readers can also show or hide specific cards using voice commands (``hide keywords'', ``show timeline'', etc.). Note that the system only supports one card of each type at a time. The user can query for a card multiple times, which will re-prompt the LLM (each card type has its own fixed prompt which the user cannot change).  }

\revision{The decision to lock content to the document was motivated by the limitations noted in prior work like \textit{Holodoc}~\cite{li2019holodoc}, where the participants expressed frustration and preferred the content to follow their views. Additionally, prior research, such as \textit{Dually Noted}~\cite{qian2022dually}, indicates that the participants preferred not to obstruct the view of the page itself. Our system displays the cards around the page and tracks them in MR with its physical movement.}

\subsubsection*{Document Tracking and Content Generation Performance}
\revision{To characterize the technical performance of our prototype, we  assessed document tracking reliability and system latency. To test tracking, we collected 60 seconds of tracking data each for 20 diverse documents (10 physical, 10 on-screen) including posters, newspapers, patents, and books---then recording what fraction of time the document was correctly registered. Tracking was largely reliable for documents with visuals (M=92\%, SD=5.4\%) but less so for text-only documents (M=64\%, SD=3.7\%) which had fewer distinct features. \change{Document tracking was more reliable when visuals (e.g., diagrams or figures) were present, but less reliable for text-only documents. Tracking performed poorly as compared to text detection because Vuforia’s image tracking depends on detecting distinctive feature points, which are more abundant in images and figures than in uniform text regions. Text-only pages often contain repetitive patterns and large blank areas harder to capture with the HoloLens camera that provide fewer robust keypoints, reducing tracking stability.} 
Performance varied with lighting (better in well-lit conditions), target size (larger images performed better), and motion (sudden movements caused temporary tracking loss).} 

\revision{We also measured the average latency of three components: AR rendering, network communication, and API responses. AR rendering averaged 100 ms on HoloLens, while network communication via socket.io averaged 40 ms, indicating responsive performance. API callbacks included GPT-4 requests (2.4 s) and our NLP module using spaCy (670 ms). Since we limited responses to a short response (150 words for a summary), GPT-4 latency remains relatively low. These latencies remained largely consistent across our two later prototypes.}


\subsection{Study Method}
Using the RS-Doc prototype, we conducted an initial usability study involving \change{12 participants (P1-P12, 6 male and 6 female, aged 19-35) recruited via snowball sampling and university-wide advertisements}. We compensated participants \$15 CAD for the one-hour session. \change{All participants reported to be proficient in English and accustomed to reading academic or professional materials. As all were enrolled university students, they met our inclusion criterion of being able to read documents of varied comprehension levels. No additional exclusion criteria were applied.} The study aimed to evaluate the system's usability, explore participants' preferences for different document enhancement strategies, and identify limitations for future iterations, with real-world use in mind. We conducted the study in a lab environment, where we first introduced participants to the HoloLens 2 and then provided an overview of the system and the set of document enhancements. \revision{Participants used our system to explore a variety of materials and media: 1)~an A4-sized magazine and an A3-sized book in physical format, and 2)~a digital document of their choice. The physical magazine and book contained the same reading content for all participants, while the digital documents were selected by participants to reflect their diverse preferences and included blog posts, course assignments, news articles, Wikipedia entries, and digital books. \change{The physical magazine and book each contained fixed content, and the same materials were used across all participants. This ensured that differences in responses could be compared consistently within each medium, without variability introduced by differing texts.} This study aimed to investigate participants' interactions and preferences across different reading materials and formats, rather than focusing on standardized tasks; therefore, we did not explicitly measure or control for difficulty levels. We tasked the participants with reading and understanding the documents provided to them, using \system{} in whichever way they saw fit and using whatever combination of enhancements best suited the document. }


Throughout the study, we encouraged participants to think aloud and documented their experiences. Following the tasks, we also conducted semi-structured interviews and a Likert scale questionnaire, including a \change{System Usability Scale (SUS) \cite{brooke1996sus}}. \change{We then performed a reflexive thematic analysis~\cite{clarke2014thematic} using transcripts of both the think-aloud sessions and interviews, from which we identified and refined key themes. }


\subsection{Results}
Overall, participants responded positively to \system{}, with most participants agreeing that the prototype was \textit{easy to use} (M=4.3/5, SD=0.98), 
\textit{easy to understand} (M=4.5/5, SD=0.67), and \textit{well-integrated with the task} (M=4.3/5, SD=0.77) on a 5-point Likert-scale questionnaire. 
Similarly, nearly all participants reported that the summary approaches provided by \system{} \textit{helped with perceived comprehension} (M=4.6/5, SD=0.514), \textit{were relevant} to the document content (M=4.7/5, SD=0.65), \change{perceived improvements in comprehension speed of the document} (M=4.4/5, SD=0.79),
and \textit{augmented their reading experience} (M=4.7/5, SD=0.65).
Finally, the System Usability Scale (SUS) returned an overall composite score of \textit{71}, suggesting a reasonable level of usability for an early-stage prototype.


\subsection{RS-Doc Observations}
Our analysis identified several themes related to participants' experience comprehending and navigating documents. We also explore participants' comparisons of the visual and text-based summarization techniques as well as their preferences for various augmentation techniques.

\subsubsection*{\revision{\textbf{Page-Level Summaries and Keyword Lists Support Quick Comprehension.}}}
\remove{\system{}'s document enhancements can significantly improve readers' comprehension. Several participants (P7, P8, P11, P12) highlighted that features such as summaries, timelines, keywords, images, and people cards provided a more comprehensive understanding of the document, allowing them to engage with the content from multiple angles and perspectives. All participants mentioned that \system{} enabled them to grasp content more quickly than with their usual methods, such as visually scanning a physical book or manually searching for timelines on a phone or PC. For instance, after reviewing the first magazine, P12 remarked, I haven't read this magazine thoroughly, but it seems these features can significantly help me understand what it's about in less time.''} \revision{Since our RS-Doc system can only capture the currently visible page, our summaries are inherently focused on page-level content rather than the entire book. Interestingly, when asked about their experiences, several participants expressed appreciation for this feature.} P5 reported, \textit{``I like the page-by-page summary—having a summary for each page does help make it feel less dense compared to a summary of the whole document}.'' Similarly, P6 emphasized, ``\textit{I like that it (the system) gives you keywords for each page and not for the whole document.}''

\subsubsection*{\revision{\textbf{Structured Overviews can Facilitate Efficient Document Navigation.}}}
Both structured overviews and in-context highlights provide useful entry points that make it \remove{easier} easy to navigate documents from a top-down perspective, even within individual pages. \remove{Broadly, participants spoke positively about how \system{} helped them navigate documents.}
Four participants (P3, P4, P6, P8) specifically mentioned using structured navigation elements like timelines to better understand the structure and flow of the text, with P4 noting that \textit{``the timeline helped me organize and navigate my thoughts very quickly while reading the text}.'' In fact, ten participants reported that the system allowed them to focus their reading by taking a ``top-down'' approach---first examining the high-level ideas and terms, then transitioning into reading the lower-level content as needed. P8 highlighted, ``\textit{[the enhancements] give you a foundation before you actually jump inside [the document].}'' This approach enabled participants to prioritize their reading, allowing them to ``\textit{focus on things that you're more unsure about}'' (P1).

\subsubsection*{\revision{\textbf{Visual aids improve engagement and memorability.}}}
Both visual and text-based enhancements offer distinct benefits to readers---visual aids provide quick, high-level overviews, while text summaries capture more complex details. Some participants preferred text-based features for gaining a deeper understanding of specific topics, while others found that visual elements improved their comprehension speed, engagement, and retention. All 12 participants rated the visual summarization techniques as helpful, with eight specifically highlighting the usefulness of images and bio cards. Several participants (P1, P4, P6, P7, P9, P10, P11) noted that visual augmentations allowed them to grasp the document’s content more quickly compared to reading text alone, with P11 stating, ``\textit{Seeing the visuals helped me grasp what's happening in the document right away}.'' Participants (P7, P8) also emphasized that visuals enhanced their engagement with the material, with P7 commenting, \textit{I'm not a big fan of reading, so having visuals helps me engage with the content}.'' Additionally, participants highlighted the potential of visuals to improve memorability, with P8 noting that for historical documents, people cards helped ``\textit{put a face to a name}.'' Similarly, P7 observed that while text-based summaries aided understanding, ``\textit{graphical elements help me remember it for a longer time}.''

\subsection{\revision{Opportunities for the Second Iteration}}
Building on these observations, we identified a number of opportunities to explore in our second iteration.

\subsubsection*{\textbf{Reader-Controllable Placement Rather Than Document Tracking}}
\revision{An intriguing finding that might contradict previous work is the role of document tracking. For on-demand document enhancement, where dedicated markers or image targets cannot be prepared in advance, unreliable or slightly jittery document tracking often leads to frustration. For example, while \textit{HoloDoc}~\cite{li2019holodoc} and \textit{Dually Noted}~\cite{qian2022dually} relied on dedicated markers or pre-defined image targets, enabling much more reliable document tracking, our workflow cannot rely on these methods. Instead, we need to rely on complex and unreliable image extraction workflows. Moreover, document tracking performed well only with visually rich documents. Participants expressed frustration when dealing with text-heavy documents, particularly under challenging conditions such as poor lighting, hand occlusion, or extreme viewing angles. Participants also expressed the need for flexibility in the placement. Therefore, while we acknowledge the benefits of document tracking, it is worth exploring reader-controlled placement as a less frustrating alternative.}

\subsubsection*{\textbf{Layouts and Visual Clutter}}
Several participants (P7, P10, P11) noted that around-document augmentation helped them focus on their reading but emphasized that the number of displayed cards and their placements should be adjusted based on the user’s needs and context. Displaying too many cards at once can lead to visual clutter and divert attention from the document itself. We observed that participants might become overly reliant on the generated content rather than engaging directly with the reading material. Whether positive or negative, 11 out of 12 participants reported that they could understand the document without actually reading it, relying entirely on the contextual information provided by RealitySummary. P9 even stated, \textit{``I didn’t even read the document, but I understand what it's about.''} Currently, the system automatically generates various types of cards when relevant, but allowing users to reduce the number of visible cards based on their preferences could help manage visual clutter and maintain focus.

\subsubsection*{\textbf{Querying and Contextual Flexibility}}
\revision{Participants highlighted the potential for enhancing the system by enabling queries on the generated summaries, which could expand its adaptability to broader use cases and contexts. The current fixed prompt approach for generating different types of cards suffices for general reading but could be further refined to address more specific reader questions. Several participants (P1, P3, P5) envisioned a more conversational system where they could ask targeted questions beyond the preset summary types. While participants found general summaries and visual information cards broadly useful and adaptable to varying contexts, they also identified opportunities to enhance aids with specific queries for timelines, comparison tables, and keyword lists that better align with their content.}

\vspace{.3cm}
Overall, the controlled nature of the study limited its ability to assess  broader applicability and real-world uses. While the initial findings are encouraging, they do not fully reflect how AR summarization and question-answering tools might perform in diverse, unstructured environments. As a result, transitioning to in-the-wild studies is essential to understanding how these augmentations might work in everyday scenarios.
\remove{ and to gain insights from prolonged, regular use in real-world settings. In conclusion, a more refined version of the system, featuring flexible augmentation placement and query-able summaries, is essential for better evaluating its effectiveness in real-world scenarios.}
\section{RS-Wild: In-the-Wild Study with the Second Prototype}


To explore these kinds of real-world use cases, we developed a second prototype (RS-Wild) that allowed us to study summarization and question-answering in-the-wild. \remove{Unlike the previous usability study, which focused on standardized reading tasks in a controlled lab setting, this study aimed to examine the versatility and applicability of our system in uncontrolled environments and with diverse reading materials. The goal was to uncover unique advantages and potential use cases for using mixed reality and AI-generated assistants in everyday settings.} 
\revision{To address performance issues with the HoloLens 2 and support more free-ranging use, we instead implemented our RS-Wild prototype for the Apple Vision Pro. This provided higher rendering quality for cards, as well as a broader field of view. 
Building on observations from our prior study, we also simplified the interface---replacing the suite of document-anchored cards from RS-Doc with a single reader-positionable card that readers could dynamically update using spoken queries.} \remove{This feature would better suit diverse applications and minimize the intrusiveness of additional content in daily use.} 

\begin{figure*}[tb]
\centering
\includegraphics[width=0.24\textwidth]{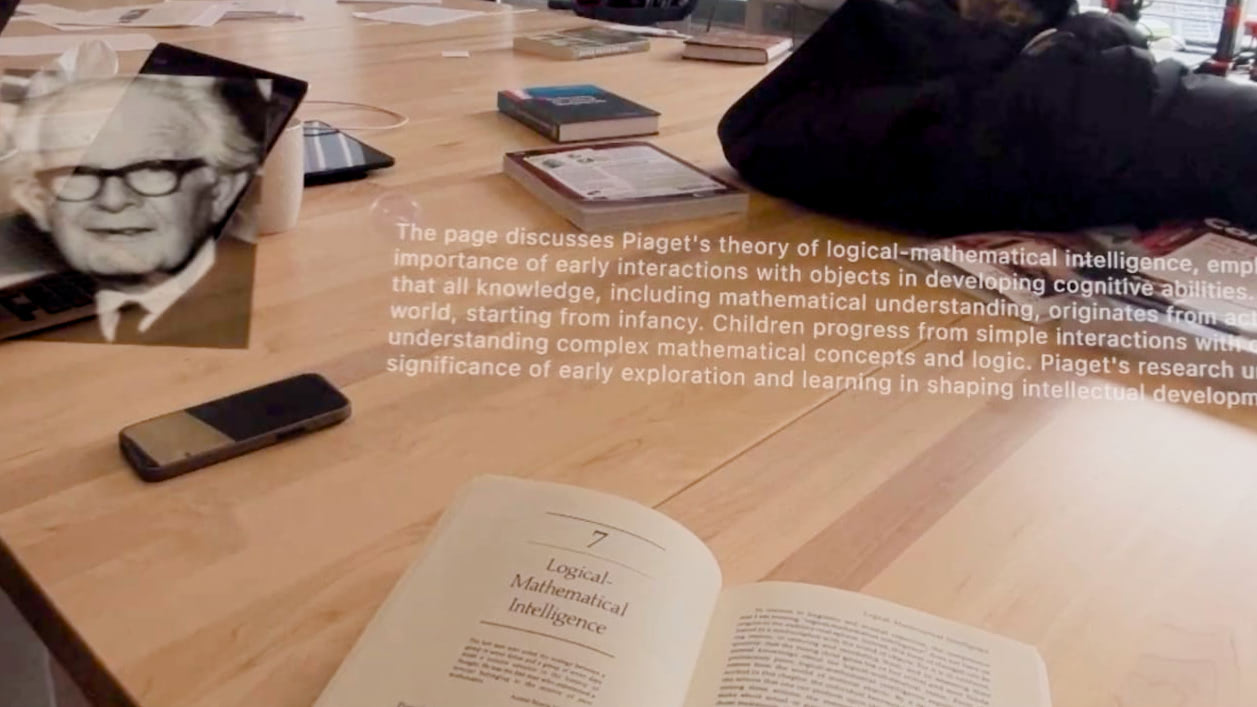}
\includegraphics[width=0.24\textwidth]{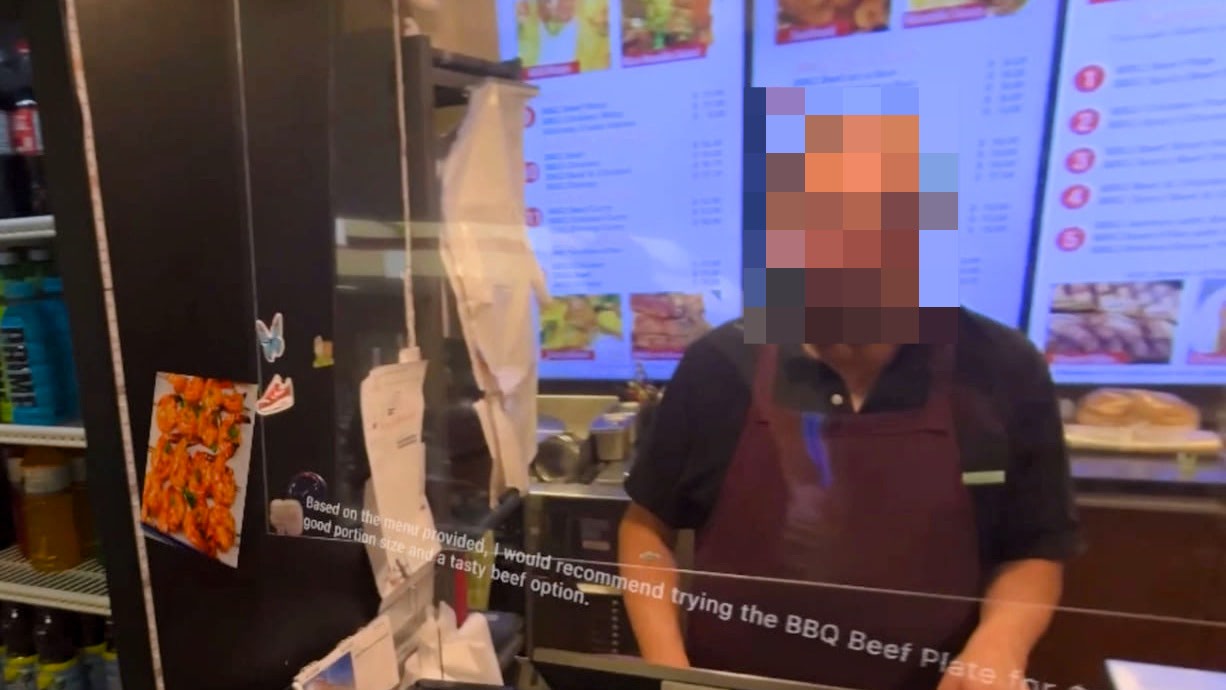}
\includegraphics[width=0.24\textwidth]{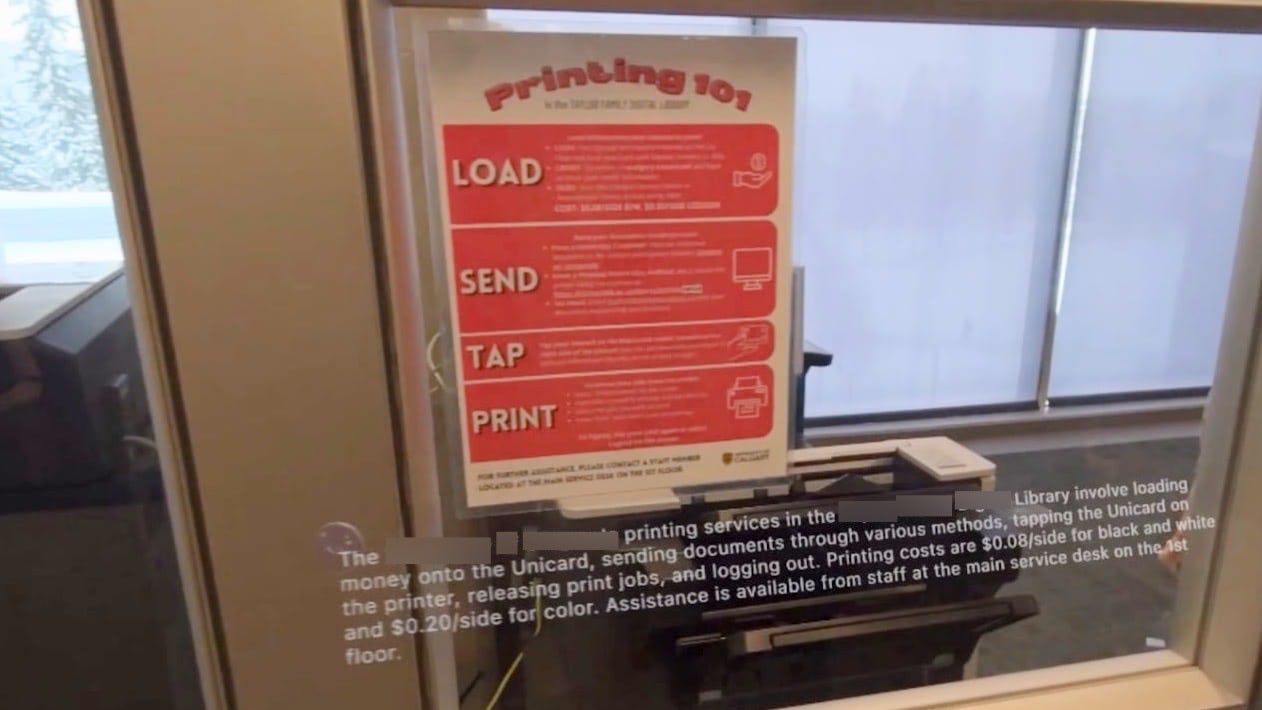}
\includegraphics[width=0.24\textwidth]{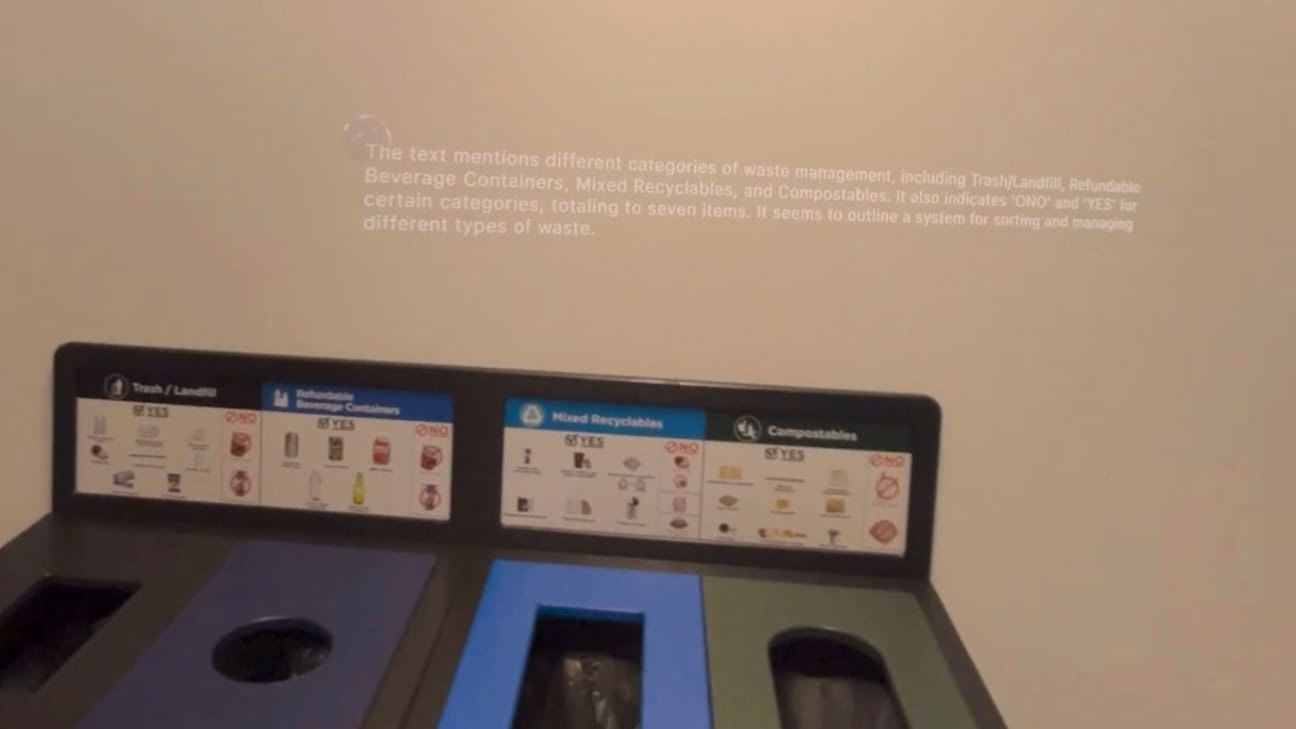}
\caption{Snapshots captured from the Vision Pro during our in-the-wild study.}
\label{fig:in-the-wild-photos2}
\end{figure*}

\subsection{Implementation}
Our RS-Wild implementation for the Vision Pro follows a similar approach to the HoloLens but with a few key adjustments. Since the Vision Pro does not yet allow direct access to the camera feed, we implemented a workaround for text extraction. First, we screencast the Vision Pro to an external iPad, which is then connected to a MacBook Air to stream the feed via QuickTime Player's mirroring feature. A web-based interface on the Mac captures the live stream from this mirror and by proxy, from the Vision Pro. Using the captured feed, we employed Google OCR to extract text, which was then processed by GPT-4 via the OpenAI API. After querying the LLM, data was transmitted from the web-based interfaces to the Vision Pro via Firebase's real-time database. 
\revision{Voice commands were handled via a separate interface (using the Web Speech API and React.js and hosted as a progressive web app) loaded on the reader's smartphone.}

\subsubsection*{Interactions} \revision{RS-Wild allows readers to request summaries and pose free-form questions about any text visible in their environment by clicking on a mic button on the smartphone interface and asking a question or clicking a ``summaries'' button. This then triggers the OCR and LLM pipeline. Responses to the most recent request are displayed on a single reader-positioned card which can be placed anywhere in the environment. This added flexibility allowed viewers to recreate any of the individual cards available in RS-Doc as well as pose new unanticipated questions and request new summary types. Using a single manually-positioned card with a transparent background also allowed readers to position that response or summary in more useful locations in their environment---placing it both close to and apart from documents as appropriate for their current task. Moreover, the user can choose to keep a history of all the text that has been captured in their session, allowing them to pose questions with a little context. Additionally, the card follows the user's position in MR space, so they always have access to it even while walking or changing locations.}

\subsection{Study Method}
\change{To explore question-answering and summarization in-the-wild we recruited a new cohort of 11 participants (W1-W11, 7 male and 4 female; ages 19-32 years) from our university through snowball sampling and university advertisements. All participants were university students accustomed to reading in academic environments, which met our inclusion criterion of being able to read documents of varied comprehension levels. No additional exclusion criteria were applied.} We compensated participants \$25 CAD for the 1.5-hour study, which included a one-hour experiment and a 30-minute interview. The participants were first called to our lab, briefed on the procedure, and given the opportunity to pick one or multiple places around the university campus (such as a library, cafeteria, store, hallway, classroom, or various labs) to explore with the system. They were allowed to freely explore any place in the campus with the headset and the study conductor would follow them. \remove{The study took place in various locations at a local university, such as a library, cafeteria, store, hallway, classroom, and multiple labs.}

Before the experiment, we collected participants' demographic information and reading habits, which we then input into the system as prior knowledge for GPT-4. Participants familiarized themselves with the Apple Vision Pro by completing the eye-tracking calibration setup. \revision{Instead of structured tasks or prepared reading materials, we encouraged participants to use the system with any text they encountered during the study, including documents, books, papers, posters, online news on mobile devices, and other materials in their environment. We also directed participants to explore various locations at a local university, rather than using spaces prearranged by the experimenters. Throughout the one-hour experiment, an experimenter accompanied the participants, observing their usage and encouraging them to think aloud about their experiences. After the experiment, we conducted semi-structured interviews, followed by a Likert scale questionnaire similar to the one used in Study 1.}


As in the previous usability study, we analyzed participants' questionnaire responses and transcripts to identify high-level trends and opportunities for augmented reading tools. \change{We performed a reflexive thematic analysis~\cite{clarke2014thematic} using transcripts of both the think-aloud sessions and interviews, from which we identified and refined key themes. }

\subsection{Results}
We gathered feedback from participants on various aspects, including the reading experience (immersiveness, understanding, distraction), perceptions of the system (content and query comprehension, trustworthiness, and privacy concerns), and overall usability (content placement and comfort using the MR headset). In general, participants reported feeling more \textit{immersed and interested} while reading (M=3.5/5, SD=1.04), and they perceived the system as \textit{effectively interpreting their queries} (M=3.3/5, SD=1.00). Their responses on the \textit{content placement} within the system were also positive overall (M=3.4/5, SD=1.29). 



Participants had varied opinions on whether the system became \textit{distracting} throughout the reading process (M=2.6/5, SD=1.37) and whether they tended to \textit{trust the accuracy and reliability} of contents provided by the system (M=2.8/5, SD=1.25). \textit{Privacy concerns} were also noted by the majority (6/11), and discomfort using the Vision Pro headset was prevalent among the respondents (8/11). 


\subsection{\textbf{\revision{RS-Wild Observations}}}
Observations from the in-the-wild deployment highlighted a wider range of use cases for augmented reading as well as some of the practical trade-offs associated with real-world use. 

\subsubsection*{\textbf{Diverse Reading Content}}
Participants engaged with a variety of books (spanning textbooks, encyclopedias, fiction, and non-fiction) as well as other types of documents (including academic papers, handouts, and class assignments) and diverse digital content accessed through smartphones and laptops (like social media posts, online news articles, and e-commerce websites). Beyond conventional reading materials, participants also used our system with everyday objects and text in the environment, such as restaurant menus, printer instructions, product nutrition labels, furniture assembly guides, posters, signage on trash bins, and advertisements. Notably, some participants even applied our system to their handwritten notes from lectures.


\subsubsection*{\textbf{Incorporating Spatial and Tangible Interactions}}
Participants appreciated the unique affordances offered by mixed reality interfaces, such as the opportunities for spatial and tangible exploration. In a library setting, for example, participants could naturally navigate through bookshelves (\textit{spatial exploration}), physically pick up a book (\textit{tangible interaction}), and browse through its pages to automatically receive summaries. 
W5 and W10 even leveraged our system to provide summaries for entire collections of books on a bookshelf, not just content from a single page. Furthermore, W1 demonstrated the ability to synthesize information across multiple books on a table, requesting summaries or connections between various books and pages. Such observations highlight how mixed reality interfaces enable spatial and tangible interactions that go beyond the limitations of traditional screen-based interfaces, which can only capture and summarize visible content.

\subsubsection*{\textbf{Creative Uses of AI-Generated Responses}}
We observed several unique and unexpected use cases that took advantage of the LLM's capacity for free-form questioning and answering. For instance, when reading a physics textbook, W8 asked the system to generate ten quizzes to help him understand quantum mechanics, transforming the textbook into a more active and personalized learning tool. Other common uses included translation (W4, W5, W6), creating learning aids (W4, W8), and generating personalized content recommendations (W5, W10). For example, W8 applied the system to his handwritten calculus notes to find answers to particular questions. Beyond traditional reading applications, the AI's capabilities supported practical decision-making. For example, W4 compared nutritional information of products on store shelves and menu items to determine healthier options, while W5 used it for guidance on waste disposal and printer operations. The LLM's ability to process a broad spectrum of inquiries clearly expanded the system's applicability and utility.

\subsubsection*{\textbf{Proactive Summaries vs \remove{On-Demand}\revision{Reader-Driven} Question Answering}}
Our system offers both 1) proactive assistance through automatic summary generation and 2) \remove{on-demand}\revision{reader-driven} assistance where readers manually ask questions. Participants shared their insights on these two functionalities by comparing their benefits and limitations. Readers appreciated how proactive summaries could often provide a useful initial overview of unfamiliar content. Preferences for proactive summaries varied with reading habits. For example, infrequent readers like W6 and W10 found these proactive summaries helpful in deciding whether to read the content at all. On the other hand, W10, who is a more frequent reader, found the constant summary updates unnecessary. In contrast, \remove{on-demand}\revision{reader-driven} question answering allowed for a more focused engagement on specific topics. For instance, W1 enjoyed ``rapid-fire'' questioning with a physics textbook, noting that it encouraged them to ask more questions than they would with traditional ChatGPT. \textit{\remove{"You have to describe it properly you have to specifically say what you're asking about what you're reading, so that kind of all of this abstracted in this system is like frees up your mind to just focus on the curiosity part Like rapid-fire questions and get rapid-fire answers".}} W2 valued how questioning was woven into reading, preserving thought flow without the disruption of toggling between reading and chat or search interfaces. Overall, participants favored the \remove{on-demand}\revision{reader-driven} mode due to its applicability to diverse situations, yet many appreciated having both options for different purposes---proactive features for implicit skimming and \remove{on-demand}\revision{reader-driven} for addressing more specific questions.

\begin{figure*}[t]
\centering
\includegraphics[width=0.24\textwidth]{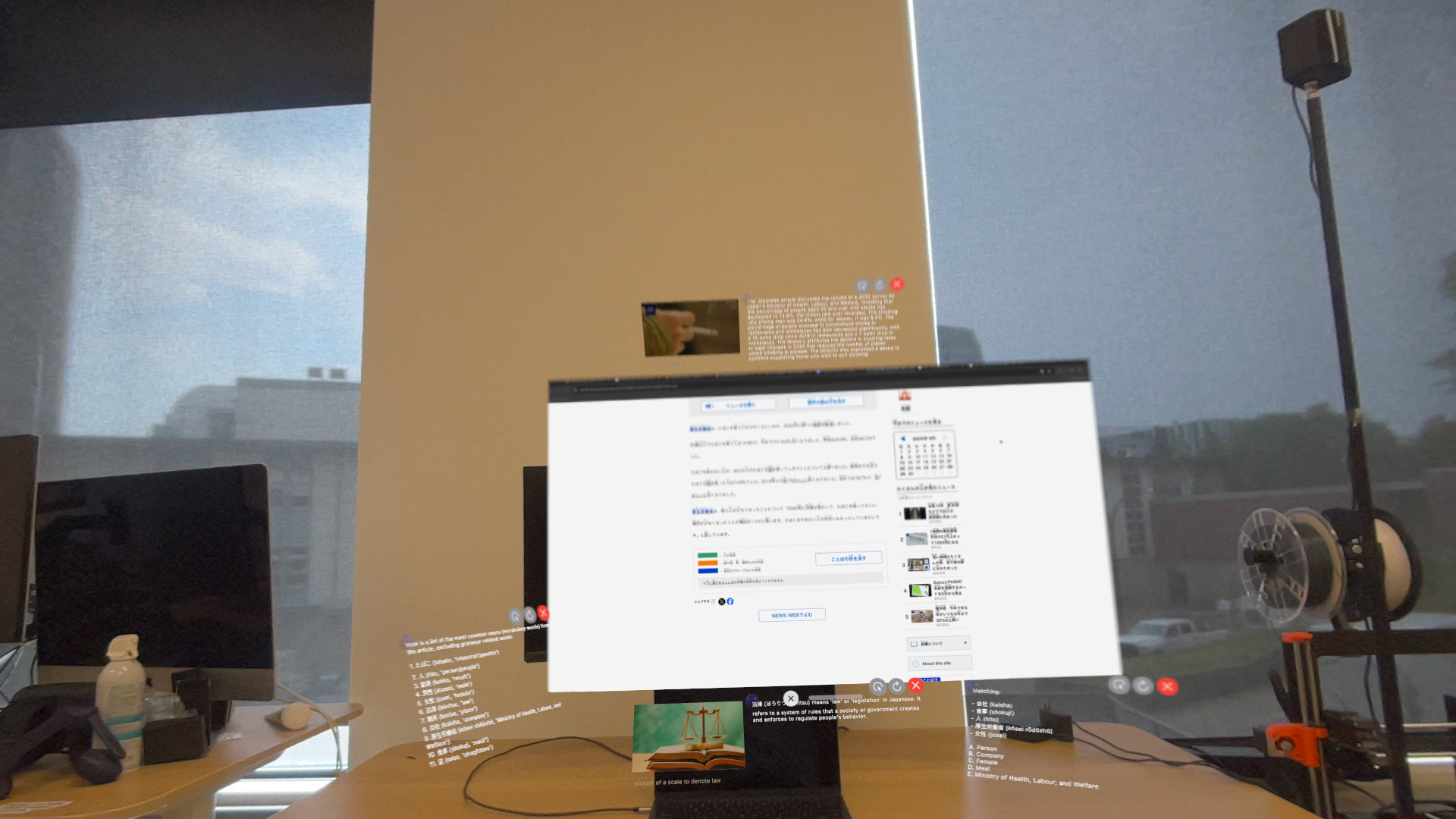}
\includegraphics[width=0.24\textwidth]{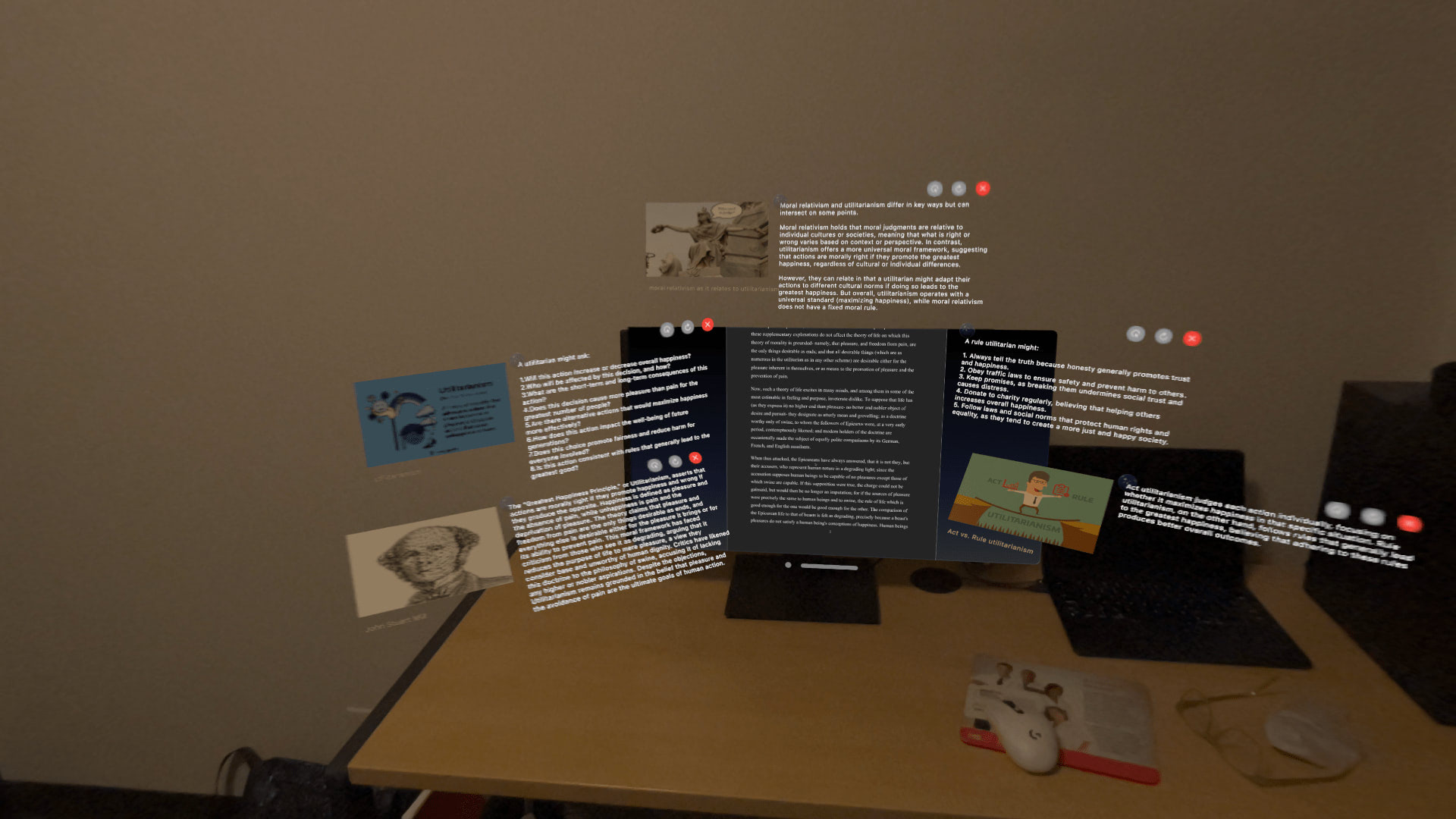}
\includegraphics[width=0.24\textwidth]{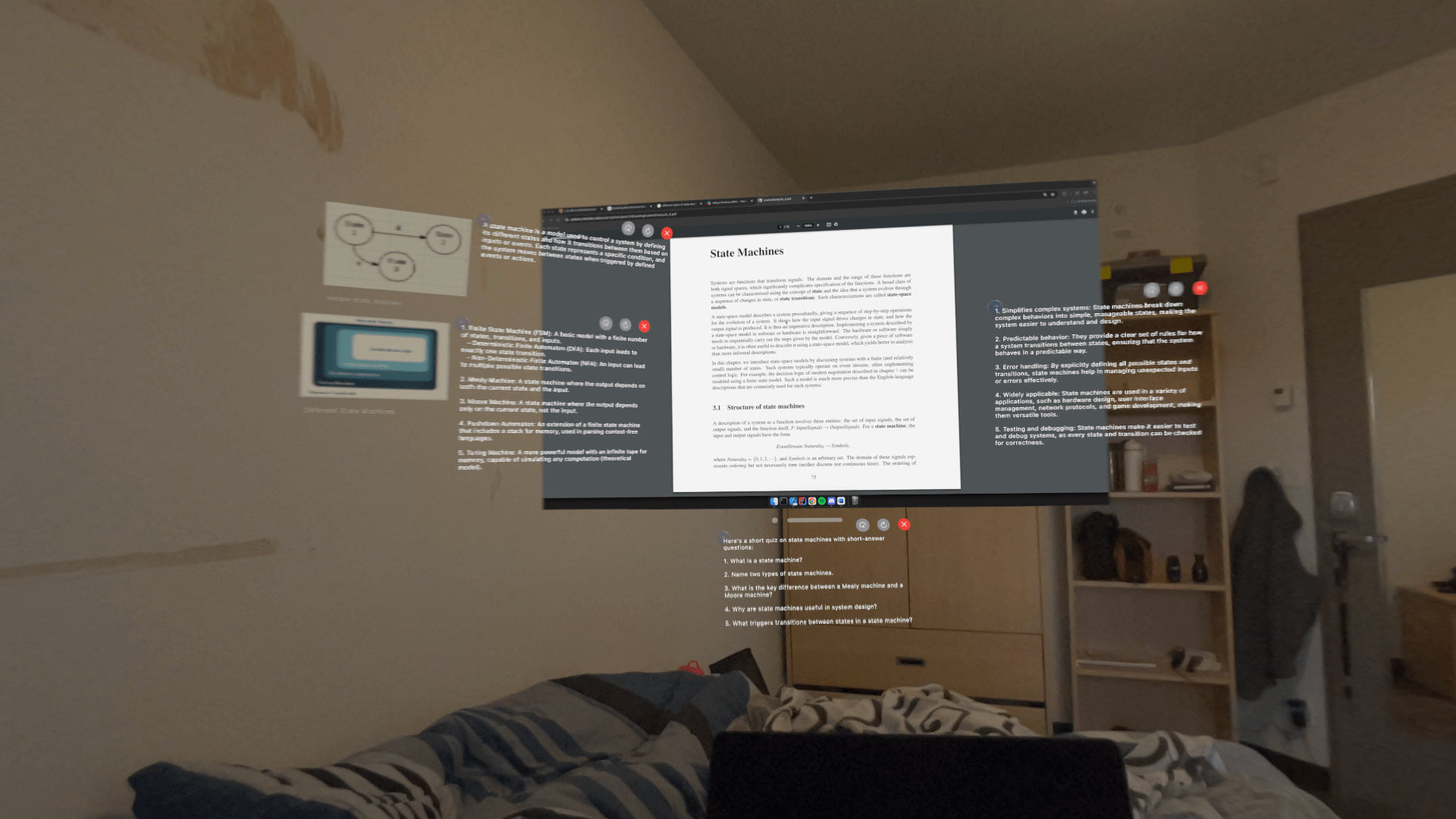}
\includegraphics[width=0.24\textwidth]{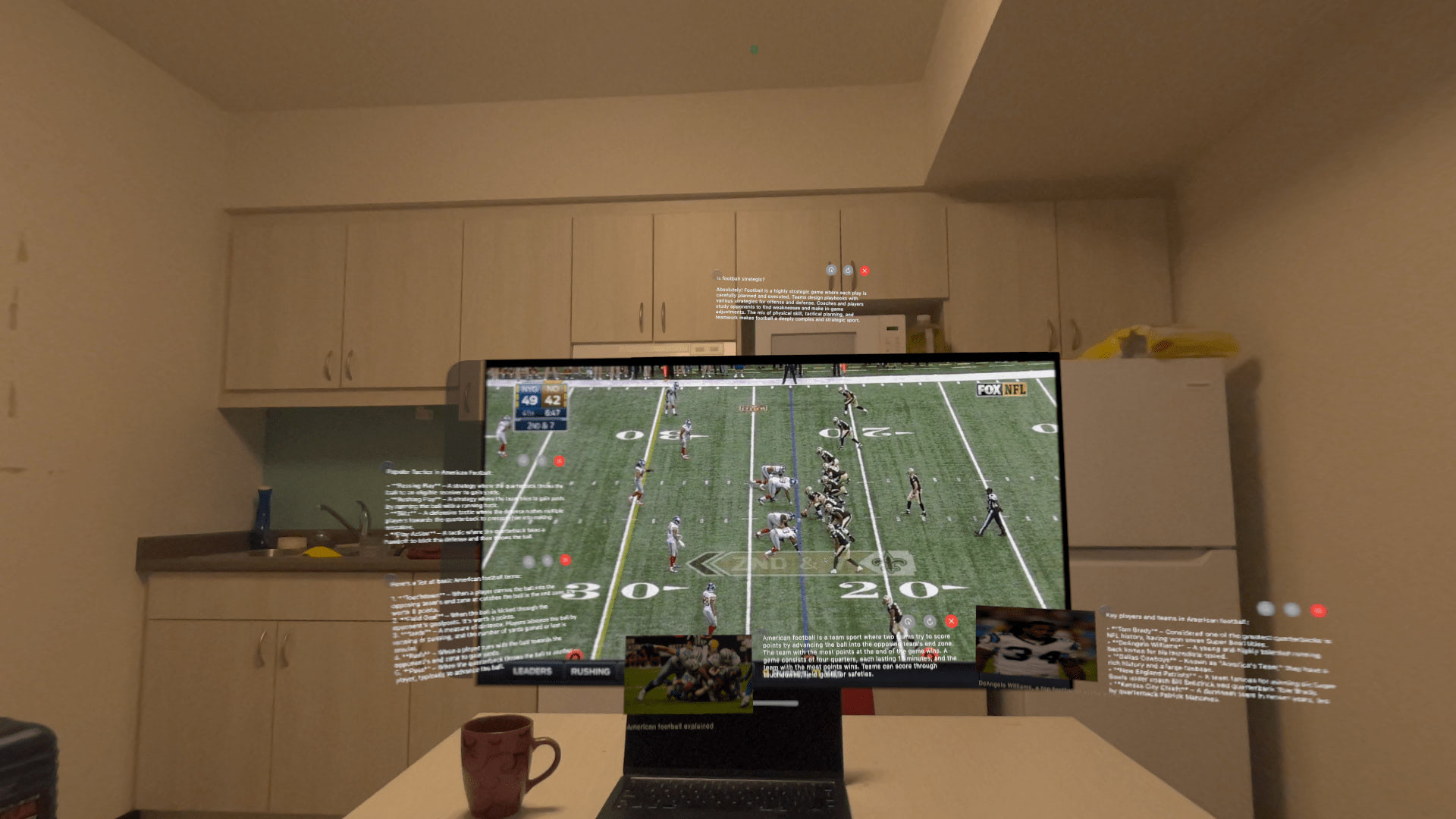}
\caption{Snapshots captured from the Vision Pro during our diary study.}
\label{fig:diary-study-photos}
\end{figure*}

\subsection{\revision{Opportunities for the Third Iteration}}
Overall, the design of the second prototype enhanced its capacity for real-world use. Compared to the earlier HoloLens 2 prototype, the Vision Pro's superior rendering quality and broader field of view provided a better reader experience. Additionally, the simple system design greatly improved usability despite the absence of document tracking. While RS-Wild showed great potential, \revision{we also identified several feature-level opportunities for improvement.}

\subsubsection*{\textbf{Sustained History and Recall}}
\revision{Participants reported that the minimalist design effectively reduced visual clutter. However, it also limited the system’s overall utility. Although readers appreciated the simple interface, many envisioned options to save or revisit previously-generated responses. Several participants suggested that a history feature or the ability to store cards for future reference would substantially enhance the reader experience. }

\subsubsection*{\textbf{Enhanced Screen Viewing for Extended Use}}
\revision{Participants highlighted the potential for improving the experience of viewing phone and computer screens through the Vision Pro, particularly for longer viewing periods. While the Vision Pro's video passthrough worked well for short-term tasks like quickly reviewing an assignment, participants envisioned improvements that could better support extended work sessions, especially when interacting with computing devices. Mirroring the reader's screen(s) visually in high resolution \textit{through} the headset would further align the system with the demands of realistic, work-related tasks in everyday contexts.}

\vspace{.3cm}
Our in-the-wild study was also inherently limited by its short-term, public use --- with each session lasting only one hour per participant. As a result, participants did not have the opportunity to fully engage with the system in their personal workflows. For instance, due to privacy concerns related to screen recording, participants rarely used our system for more personalized content, such as emails, messages, or calendars. Furthermore, since we encouraged participants to explore multiple scenarios, they tended to experiment broadly rather than focusing on any specific task in depth. While the in-the-wild study provided valuable insights into real-world use cases, it only captured short-term interactions in public spaces. There is still much to learn about the sustained, regular use of reading assistants in daily life, particularly in how readers interact with information over time, revisit previous queries, or support ongoing tasks.

\section{RS-Diary: Diary Study with the Third Prototype}

To explore more realistic longer-term use cases, we developed a third prototype variant (RS-Diary) incorporating feedback from the in-the-wild study. This version (\autoref{fig:diary-study-photos}) introduced the ability to create, retain, and organize multiple information cards, along with support for mirroring content from external devices directly to the Vision Pro.

\subsection{Implementation}
RS-Diary builds on the earlier Vision Pro implementation with two key improvements: \revision{1) Readers can now add multiple draggable cards on-demand by either making a new query or clicking an ``add'' button (which spawns a new summary card based on the document content currently in front of them). They can also spatially anchor these cards anywhere in their immediate environment and can re-select existing cards to update them with new prompts or text content.} These cards are stored as JSON objects in a real-time database. 2) The system also allows readers to mirror their device screen to the Vision Pro for high-resolution viewing. This functionality---implemented through a WebRTC video call or via Mac Virtual Display on the Vision Pro---lets readers view their computer or tablet screens at full resolution in the MR environment and also allows RS-Diary to more accurately capture and process on-screen content. \revision{This increased fidelity allows readers to perform a wide variety of everyday reading, coding, browsing, and watching activities for extended periods without the legibility and eye strain experienced with video passthrough}.

\subsection{Study Method}
Four authors and collaborators (AA, AM, AJ, AW, and CT \footnote{The first, fourth, fifth, and sixth authors of this submission and one external acknowledged collaborator. The second letter of each code corresponds to the author/collaborator's first initial.}) used the system over the course of 3 weeks, integrating it into their day-to-day activities and accumulating 30 total hours of use. Due to physical fatigue and battery limitations of the Vision Pro, each session lasted less than 2 hours. \change{The participants independently documented their use cases, insights, frustrations, thoughts, and reflections while using RS-Diary. During the study, each participant recorded their observations via a smartphone audio recorder and a personal reflections diary for notes, following a think-aloud process. Additionally, the first author conducted reflective debrief conversations with the other four participants at the end of the study to gain deeper insights into their experiences.} \change{We then conducted a reflexive thematic analysis~\cite{clarke2014thematic} using transcripts of these discussions, from which we identified and refined key observations.}


\subsection{Results}
Throughout the diary study, participants used the system in various settings, including labs, kitchens, outdoor areas, and offices, though the majority of usage took place at desks in office environments. They engaged with a wide range of reading materials, such as academic papers, news articles, code, emails, calendars, Google Scholar, classroom assignments, lecture slides, textbooks, and YouTube videos. Participants quickly adapted the system to fit a variety of personal, academic, and technical tasks. For example, AW used the assistant not only to summarize email content but also to gather higher-level insights, such as identifying how many emails were sent by a particular sender. AM utilized the system to support programming assignments, where they used the system to break down complex coding problems into manageable steps and provided detailed explanations of specific code snippets. AM also leveraged the system to create learning aids, such as generating to-do lists and quiz questions based on problem descriptions. The ability to spatially organize tasks and concepts allowed him to balance his workload, maintain focus, and dynamically track his progress. Additionally, in his language learning efforts, AM used the assistant to generate contextual vocabulary lists from Japanese reading materials, which helped him focus on the most relevant and frequently used words. CT employed it to extract a step-by-step sequence of keyboard commands from an article on rebooting a MacBook. Overall, the diary study highlighted the system's flexibility in supporting diverse tasks, from email management and coding to technical troubleshooting, language learning, and academic research. 

\subsection{\revision{RS-Diary Observations}} 
From these experiences, participants shared various thoughts, frustrations, and potential future improvements. Our diary study also provided insight into the use of an always-on reading assistant in more realistic work-oriented settings.

\subsubsection*{\textbf{\revision{Quick Summary Responses Meet Many Immediate Needs}}}
Participants valued the system's ability to deliver quick, concise responses, which were often sufficient for immediate clarifications. AM, for instance, appreciated how the system provided brief answers when they needed a rapid comparison of philosophical concepts, like distinguishing between act and rule utilitarianism. In many cases, just a high-level summary of the page in view was enough to satisfy their needs. AW found this method to be a useful shortcut, as pressing the virtual ``add'' button was faster and less disruptive than speaking and often provided exactly the information they were seeking. These simple, proactive summaries were particularly beneficial for tasks that required quick comprehension and minimal interruption.

\subsubsection*{\textbf{Using Generated Cards as Peripheral References}}
\revision{Participants highlighted the advantages of cards that could exist in the periphery of their current workspace.} For instance, AM used the system as a reading aid while learning Japanese. Instead of frequently opening a dictionary, which disrupted the reading flow, they generated vocabulary flashcards around the document. This provided language support without interrupting their reading and learning process. AW also noted that they would often cluster and group these generated cards thematically to the sides of the display for quick reference and then delete them once they were no longer needed. Similarly, AA used the system to ``pin'' generated code versions and key concepts around their workspace, allowing for easy access without breaking focus and helping them quickly revisit important points as needed.

\subsubsection*{\textbf{Spatial Organization for Recall and Multitasking}}
Perhaps the biggest emergent pattern in the study was how all participants leveraged the system’s spatial organization features, similar to arranging post-it notes on a board. Both AA and AM utilized this capability to group related results and summaries together. For example, AM clustered cards on CPU architecture, placing related topics like buses and registers nearby, much like organizing post-its by theme. This spatial arrangement enabled quick access and improved recall, as readers could visually track where they had placed specific information. Similarly, AA pinned cards documenting different problem-solving methods across their workspace, ensuring easy access when needed. These organizational structures supported memory retention and task management by allowing readers to arrange information in a way that aligned with their thought processes.



\section{Lessons Learned from Three Studies}

Drawing on results from across our prototyping and evaluations, we close by showcasing the unique benefits of combining MR and AI tools, enumerating potential use cases, and highlighting opportunities for future research.


\subsection{\revision{Key Insights from the Three Studies}}

\subsubsection*{\textbf{Minimal Context Switching Enhances Focus and Flow}}
Participants across all three studies appreciated the integration of LLM into the MR interfaces, in part, because it helped them minimize context switching, allowing them to maintain greater focus. Several participants reported that the system’s automatic text extraction feature helped them stay in a flow state. For example, AM noted that the interface allowed them to concentrate on their work without the mental disruption of switching between tasks or manually inputting context, even when using a computer screen. In particular, members of the author team in the RS-Diary study remarked on how the seamless workflow kept them engaged, enhancing productivity and reducing cognitive load. AA highlighted how the system functioned like a ``teacher'' by offering real-time assistance without requiring them to step away from their tasks. 
Participants also remarked that on-demand support boosted their efficiency during complex tasks like coding or writing, where focus is critical. 

\revision{A key benefit of this workflow is its support for \textit{implicit inputs}, which dramatically lower the barrier to asking questions of LLMs. This automatic, continuous content capture broadens the range of applications beyond what participants initially envisioned. For instance, W4 mentioned they had never thought to type a question into ChatGPT when looking at a product label. Even for common reading tasks, they appreciated that the summary was automatically generated and updated in the background as they flipped through pages in a library. Overall, the always-on, implicit assistant made it effortless to switch context and maintain focus.}

\subsubsection*{\textbf{Potential Risks of AI-Driven Augmented Reading}}
Participants also expressed several concerns about potential risks, notably privacy issues due to the possibility of constant surveillance. Despite this concern, most participants expressed interest in continue to use these kinds of systems, provided they could control when it was active. For example, W1 expressed the desire to capture all text throughout her entire life, allowing for an extensive searchable archive of everything she has encountered. Such use cases, however, also pose social privacy questions---particularly when systems might record other individuals or their content. 

Additionally, participants mentioned the risk of decreased motivation and skill development. For example, W6 mentioned that relying too heavily on the system for translations could reduce the motivation to learn new languages. Similarly, W11 worried that immediate access to answers for every question might contribute to a broader intellectual complacency. Overall, participants wanted the flexibility to choose when and how the system would be used to mitigate these risks.

\subsubsection*{\textbf{Trust of AI-Generated Content}}
Recognizing the known shortcomings or current LLMs, participants also approached the system's responses with caution, often verifying the answers by reading the source materials and highlighting the necessity of consulting additional sources. There was a general agreement among participants in all studies that they would not fully trust the system's responses to be complete or correct. 
Conversely, we noted some intriguing discussions regarding how participants perceived our system, either as a mere tool or more like a personal companion. While most participants engaged with the system as a conventional chatbot, some regarded it more as a personal companion than a mere tool. Notably, W3 and W5 described the system as a friend for discussing and inquiring about their readings. The system's ability to provide always-on, immediate responses facilitates this personification. 

\subsection{Unique Benefits of Combining MR and AI} 
We identified two unique benefits of combining MR with AI: 1)~implicit input and 2)~peripheral output.

\subsubsection*{\textbf{Always-On Implicit Input}} 
The always-on camera in MR enables \textit{implicit} input, distinguishing it from the explicit input required in desktop interfaces, such as opening an app, typing queries, or copy-pasting contextual text. Participants in in-the-wild and diary studies appreciated the system's ability to capture information without requiring explicit effort. This always-on input allows for hands-free exploration in both spatial and temporal contexts, such as scanning entire bookshelves as a query or utilizing the history of flipped pages. Implicit input reduces cognitive load and minimizes context switching, as readers are not required to actively specify what information should be used for the LLM’s context.

\subsubsection*{\textbf{Peripheral and Background Output}}
We found that MR’s spatial output allowed information to be presented in a peripheral, background manner. In contrast, other output modalities, such as audio or screen-based visuals, often require the reader’s foreground attention to consume generated content, disrupting focus and interrupting workflows. With MR and peripheral outputs, however, AI-generated content becomes less intrusive, as it can be easily ignored if not needed, simply by choosing not to look at it. For example, participants in the diary study often appreciated this feature for learning aids, such as understanding textbooks or accessing always-on vocabulary references when learning a foreign language. This peripheral output is a unique aspect of the integration of MR and AI. We believe this capability would promote more \textit{calm} human-AI interactions in the future, reducing disruptions to focused attention by combining implicit input with unobtrusive background support.


\subsection{Emergent Use Cases} 
Both the in-the-wild and diary studies revealed a variety of interesting emergent use cases for MR text augmentation. To give a more comprehensive view of how readers might use these systems, we identify seven distinct approaches:

\subsubsection*{\textbf{Summarizing Content}}
The primary use case was to summarize content. This was the most commonly observed behavior across various contexts. Participants summarized a wide range of texts, including research papers, blog posts, news articles, books, and instructional guides.

\subsubsection*{\textbf{Extracting Information}}
Participants also utilized the system to extract key information, which can be translated into more consumable formats like tables of contents, keyword lists, timelines, and comparison tables. For example, one participant extracted vocabulary flashcards to assist with foreign language reading, while another participant created a table of keyboard shortcut commands from a troubleshooting guide to use as a reference. In these examples, the consumable format, such as a table or vocabulary list, serves as helpful references for their tasks.

\subsubsection*{\textbf{Querying External Sources}}
Participants frequently used extracted text as a query to retrieve additional information from external sources. Participants often queried specific names or keywords from their reading materials. For instance, one participant retrieved details about players, teams, and standings from a sports-related article, while another fact-checked claims on-demand when reading news articles.

\subsubsection*{\textbf{Asking for Suggestions}}
Participants frequently asked for recommendations or suggestions based on the current context---for instance,  asking for book recommendations when looking at a bookshelf. Outside the context of reading, one participant asked for advice on which trash bin to use based on the disposal instructions on an item, while others inquired about healthier menu options or products. They also asked for recommendations from external sources, including asking for related research papers when reading academic articles. 

\subsubsection*{\textbf{Aggregating Information}}
An interesting use case involved using the system to aggregate high-level information. Unlike simple information extraction, aggregation often involves operations like filtering and counting. For example, some participants used the system to count emails based on specific filters, while others filtered and counted first-author papers from a researcher's Google Scholar page. This allowed readers to perform quick, approximate queries that would normally require more complex searches.

\subsubsection*{\textbf{Transforming Content}}
Participants also used the system to transform their reading content into interactive, actionable outputs. For instance, some participants applied our system to generate quizzes from class notes or textbooks, while others decomposed high-level assignments into step-by-step to-do lists. One participant asked the system to create reminder cards that remained spatially persistent. Participants also used our system to translate from one language to another. Unlike information extraction, these tasks involved converting content into actionable, interactive outputs.

\subsubsection*{\textbf{Generating New Content}}
Finally, participants generated new content based on existing text. In one example, a participant requested multiple variations of a written paragraph, which they then organized spatially. Others used the system to create code based on instructions or to debug existing code snippets.

\section{Conclusion}
\change{RealitySummary is the first exploration of on-demand MR reading assistants that combines OCR and LLM to generate spatially-situated on-demand reading aids for both physical and digital documents. Through our empirical studies, we highlighted new possibilities for on-demand, AI-powered reading support. This paper presented an iterative development of the system, along with key findings and lessons learned from three iterations and studies: Study 1 surfaced usability concerns and reader perceptions in controlled settings, Study 2 revealed challenges and opportunities of in-the-wild deployment, and Study 3 shed light on longer-term adaptation through a diary study. Together, these studies provide a more complete picture of how MR reading assistants can complement everyday reading. At the same time, our work faces limitations. The participant pool was small, with participants being university students. The in-the-wild deployment was relatively short in duration, and system performance was constrained by the practical limitations of the Apple Vision Pro, including its form factor and battery life. Additionally, Study 3 relied on author-participants, which may introduce bias; however, our reflective diary study approach provided valuable long-term insights that would have been difficult to obtain otherwise. These limitations point to opportunities for future work: evaluating the system with larger and more diverse populations, comparing always-on versus on-demand support, and extending deployments to longer-term contexts. We hope our work will inspire the HCI community to further explore the potential of integrating MR and AI in everyday reading applications.}

\begin{acks}
\change{This research was partially funded by the NSERC Discovery Grant RGPIN-2021-02857, JST PRESTO Grant Number JPMJPR23I5, the Canada Research Chairs Program, and an Adobe Collaborative Research Gift. We also thank all of our study participants. We would like to acknowledge Tafreed Ahmad for helping with the studies and giving us constructive feedback.}
\end{acks}

\balance
\bibliographystyle{ACM-Reference-Format}
\bibliography{references}

\end{document}